\documentclass[11pt]{iopart}
\pdfoutput=1
\expandafter\let\csname equation*\endcsname\relax
\expandafter\let\csname endequation*\endcsname\relax
\usepackage[numbers,sort&compress]{natbib} 
\usepackage[pdftex]{graphicx,color}
\usepackage{amsmath,amsfonts,amssymb,wasysym,graphicx,capt-of,ifthen,calc}
\usepackage{enumerate,float,url,color,subfigure}
\usepackage{amsmath}
\usepackage{graphics}  
\usepackage{psfrag} 
\usepackage[latin1]{inputenc}
\usepackage{color}
\usepackage{rotating}
\usepackage{url}
\usepackage[colorlinks=true, urlcolor=blue, anchorcolor=blue, citecolor=blue,filecolor=blue,linkcolor=blue,menucolor=blue]{hyperref}

\begin{document}
\title{Birds on a Wire}
\author{P. L. Krapivsky}
\address{Department of Physics, Boston University, Boston, MA 02215, USA}
\address{Santa Fe Institute, 1399 Hyde Park Road, Santa Fe, NM 87501, USA}
\author{S. Redner}
\address{Santa Fe Institute, 1399 Hyde Park Road, Santa Fe, NM 87501, USA}

\begin{abstract}
  We investigate the occupancy statistics of birds on a wire. Birds land one by one on
  a wire and rest where they land.  Whenever a newly arriving bird lands
  within a fixed distance of already resting birds, these resting birds
  immediately fly away.  We determine the steady-state occupancy of the wire,
  the distribution of gaps between neighboring birds, and other basic
  statistical features of this process.  We briefly discuss conjectures for
  corresponding observables in higher dimensions.
\end{abstract}


\section{Introduction}
\begin{figure}[ht]
  \centerline{
\includegraphics[width=0.475\textwidth]{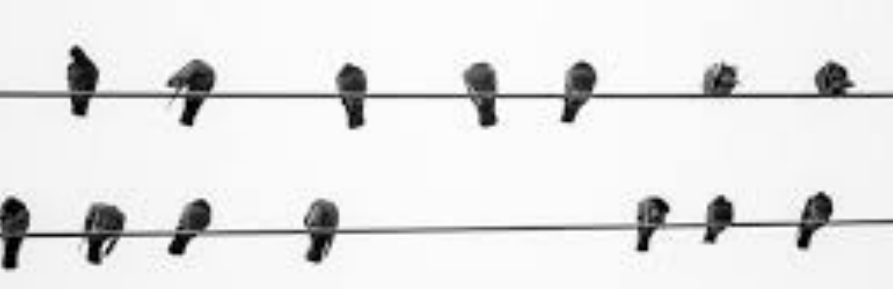}  }
  \caption{Birds on wires.}
\label{fig:wire}  
\end{figure}

Statistical mechanics provides us with the ``eyes'' to appreciate
collective phenomena in quantitative and insightful ways.
Figure~\ref{fig:wire} illustrates this synergy between phenomenology
and analysis: birds alight one at a time to rest at random positions
on a wire.  We postulate that birds are sociable but skittish---if a
newly arriving bird lands within a specified distance of any resting
birds, they immediately fly away. A first question to address is: What
is the dynamics of this process?  Eventually, a steady state is
reached in which the average arrival and departure rates are equal and
this prompts several questions.  For example, What is the steady-state density
of birds on the wire?  What are the separations between adjacent birds? 

While there has been much research on the spatial patterns of moving
animal
groups~\cite{vicsek1995novel,toner1995long,toner1998flocks,couzin2002collective,couzin2005effective,ballerini2008interaction,ballerini2008empirical,vicsek2012collective},
the spatial organization of static groups seems less studied (see,
however, \cite{vseba2009parking,aydougdu2017modeling}).  We formulate
the ``pushy birds'' (PB) model (see Fig.~\ref{fig:processes}) to mimic
the spatial organization that results from repeated landings and
departures of birds.  This idealized model is similar in spirit to
models of flocking and
schooling~\cite{vicsek1995novel,toner1995long,toner1998flocks,couzin2002collective,couzin2005effective,ballerini2008interaction,ballerini2008empirical,vicsek2012collective}.
While our model focuses on the one-dimensional geometry with local
interactions, it naturally extends to longer-range interactions that
may lead to self-organized cooperative behavior, as in forest-fire
models~\cite{FF-Bak90,drossel1992self,drossel1993exact,FF-Maya93,FF-Grassberger02,FF-Jensen20}.
A generalization to higher dimensions leads to a dynamic version of
the famous sphere packing problems in arbitrary dimensions (see,
e.g.,~\cite{Rogers,Levenshtein79,Odlyzko79,Conway,Hales05,Aste,Cohn16,Cohn17})
for which little is known.

Our PB model also resembles random sequential adsorption
(RSA)~\cite{Flory39,Evans93,talbot2000car,torquato2002random,adamczyk2017particles,KRB,Itoh,Newby13,krapivsky2020large},
where fixed-shape particles impinge on open regions of a substrate and
stick irreversibly.  One example of RSA that is close to our PB model
is the ``unfriendly seating arrangement''
problem~\cite{Shepp62,friedman1964problem}, where people arrive one at a time
at a luncheonette and and sit at a counter.  People are all mutually
unfriendly so they choose seats at random but never next to another
person.  The luncheonette reaches a static jammed state of density
$\rho_\text{jam} = \tfrac{1}{2}(1-e^{-2})\approx 0.432$, after which
additional patrons cannot be accommodated.  In contrast, our PB model
reaches a steady state that is constantly changing locally, but its
global properties are stationary and independent of the initial
conditions.

While our model is couched in terms of birds, it should not be taken
literally as a description of real birds.  There are many other
influences that the determine how birds organize themselves on a
spatially restricted landing spot, such as a wire. Nevertheless, the
behavior of our admittedly unrealistic model is rich and perhaps this
study provides some initial steps to understand the organizational
dynamics of more realistic models of the arrival and departure of
birds at some resting spot.  We view our PB model has being akin to
some of the idealized forest-fire models that were proposed long ago
in the statistical physics
literature~\cite{FF-Bak90,drossel1992self,drossel1993exact,FF-Maya93}.
These abstract models miss many features of real forest fires;
nevertheless, the phenomenology that arises from this class of models
is extremely rich and led to many advances about self-organized
criticality~\cite{bak1987self}.  It is in this impressionistic spirit
that we investigate our PB model.

\section{One-Dimensional Lattice}

It is conceptually simplest to formulate a discrete version of our PB
model in which birds land on empty sites of a one-dimensional lattice;
we later treat a continuous version.  Each landing event of a bird
scares away birds on adjacent lattice sites (if they are present) so
that they fly away.  Our analysis of the PB model focuses on $V_k$,
the density of voids of length $k$.  A void of length $k$ is defined
as the following arrangement of birds and vacancies
\begin{equation*}
\circ\bullet\underbrace{\circ\,\ldots\,\circ}_k\bullet\,\circ\,,
\end{equation*}
where an occupied site is denoted by $\bullet$ and an empty site by $\circ$.
Since birds cannot be adjacent, the sites next to each bird outside any void
must also be empty.

\subsection{The void densities}

\begin{figure}[ht]
\centerline{\subfigure[]{\includegraphics[width=0.17\textwidth]{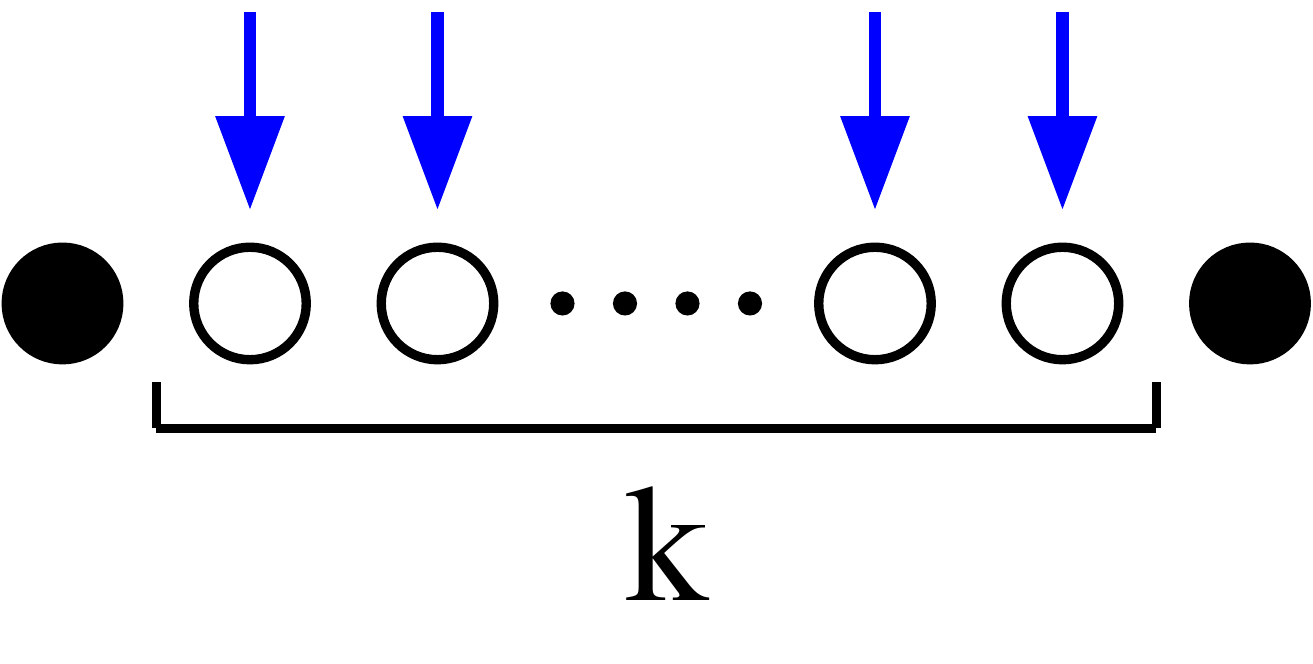}}~~~~~
\subfigure[]{\includegraphics[width=0.21\textwidth]{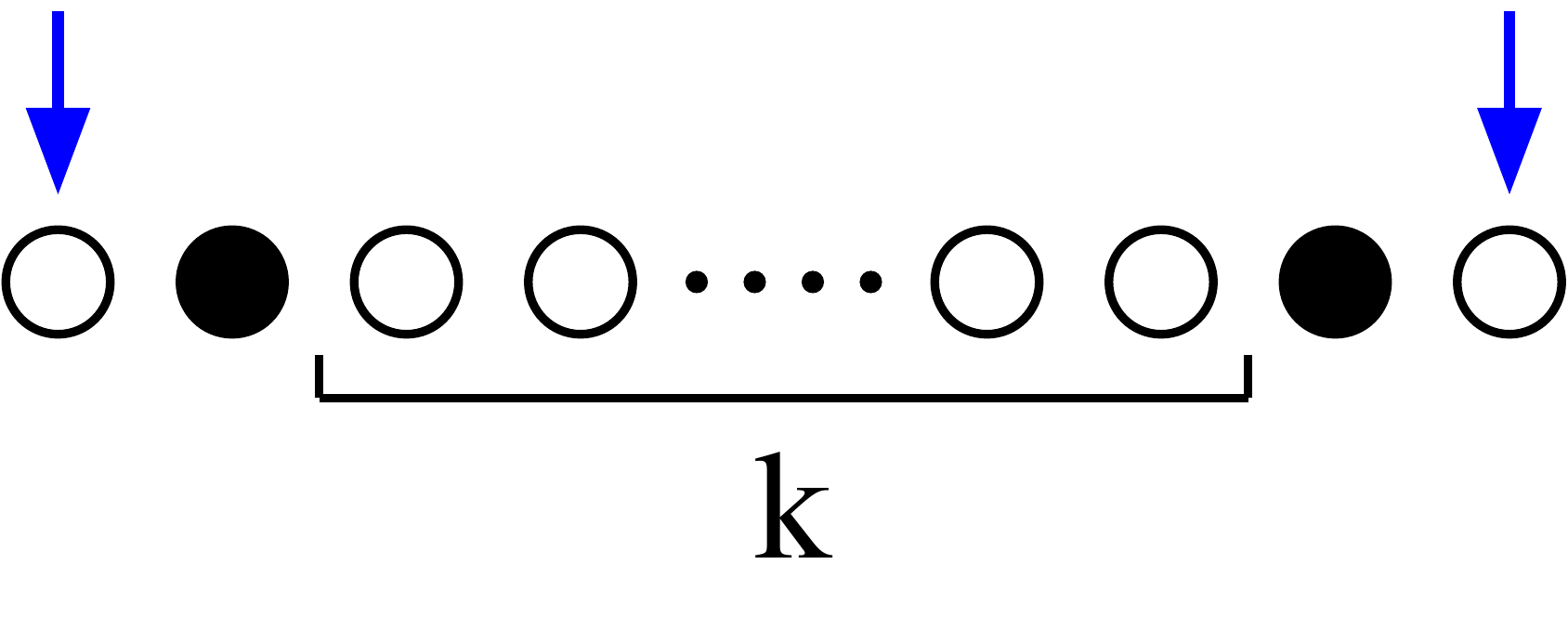}}~~~~~
\subfigure[]{\includegraphics[width=0.2\textwidth]{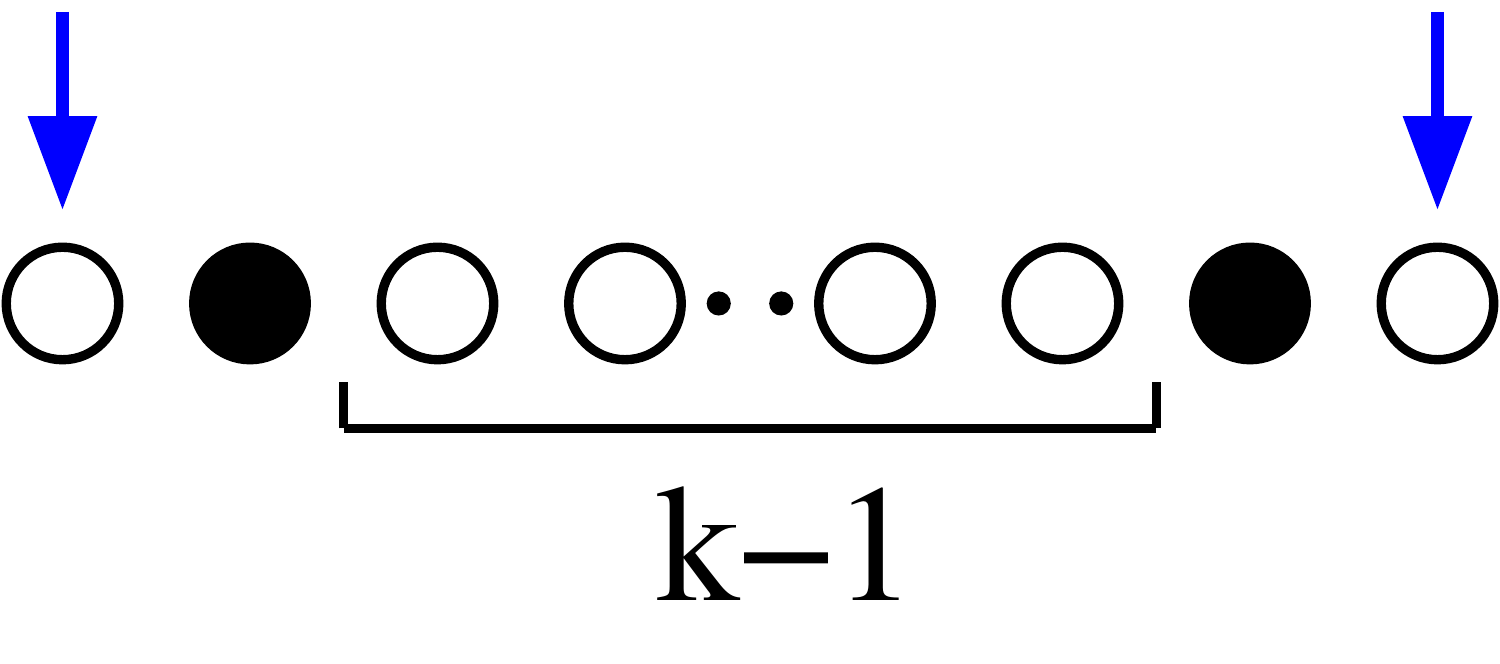}}~~~~~
\subfigure[]{\includegraphics[width=0.24\textwidth]{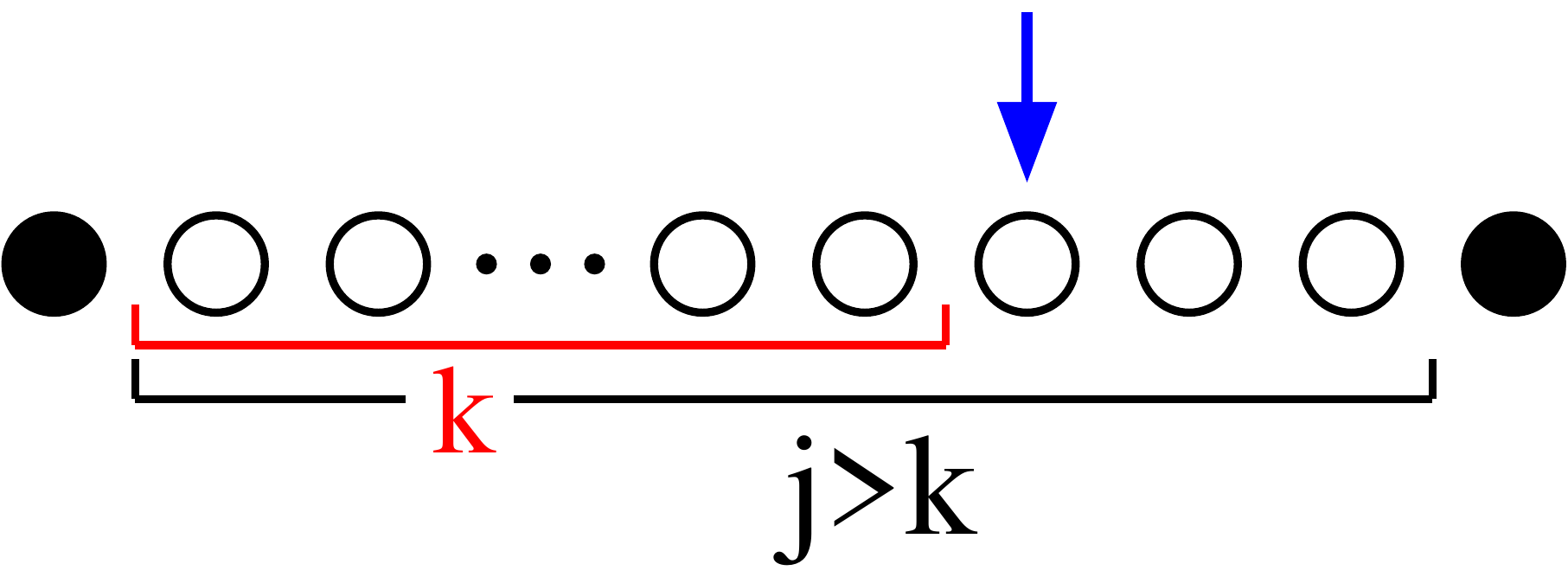}}}
\caption{Processes that contribute to changes in the void densities in
  the PB model of Eq.~\eqref{evol}.  The vertical arrows indicate the
  possible locations for a bird to land.  In (d) only one of the two
  possible landing spots that creates a void of length $k$ is shown.}
\label{fig:processes}  
\end{figure}

The void densities change in time according to the following rate equations:
\begin{equation}
\label{evol}
\dot V_k = -kV_k - 2V_k+2V_{k-1} + 2\sum_{j\geq k+1}V_j
= -(4+k)V_k+2\sum_{j\geq k-1} V_j\,,
\end{equation}
where the overdot denotes time derivative.  Each of the terms on the right
corresponds to one of the processes shown in Fig.~\ref{fig:processes}.  The
first term accounts for the loss of a $k$-void due to a bird landing anywhere
within this void (Fig.~\ref{fig:processes}(a)).  The second term accounts for
the loss of the $k$-void when a bird lands in either of the two sites just
outside this void.  Immediately afterwards, the adjacent bird at the edge of
the $k$-void flies away, so that a $k$-void disappears
(Fig.~\ref{fig:processes}(b)).  The third term accounts for the gain of a
$k$-void when a bird lands on either of the two sites just outside a void of
length $k-1$; this ultimately causes an increase in the number of $k$-voids
(Fig.~\ref{fig:processes}(c)).  The last term accounts for the gain of
$k$-voids when a bird lands within a $j$-void, with $j>k$, such that a
$k$-void is created.  If $j\ne 2k+1$, there are two possible landing sites
(Fig.~\ref{fig:processes}(d)), each of which creates one $k$-void.  If
$j=2k+1$, there is a unique landing site in the middle of the $j$-void that
creates two $k$-voids.

The void distribution also satisfy the following basic conditions that will
be useful in solving the model:
\begin{equation}
\label{conditions}
V_0 = 0\,, \qquad\sum_{k\geq 0}V_k= \rho \,, \qquad  \sum_{k\geq 0}(k+1)V_k=1.
\end{equation}
The first equality states that voids of length 0 cannot exist because this
corresponds to two birds being adjacent.  The one-to-one correspondence
between each void and exactly one bird leads to the second equality between
void densities $V_k$ and the overall density $\rho$.  The last equality
states that the length of all voids plus the bird at one end of each void
equals the total length.

Summing Eqs.~\eqref{evol} over all $k\geq 1$ and using the sum rules
\eqref{conditions}, we obtain a closed equation for the
density, $\dot\rho = 1 - 3\rho$.  For an initially empty system, the solution is
\begin{equation}
\label{rho:sol}
\rho = \frac{1}{3}(1-e^{-3t})\,.
\end{equation}
Thus the approach to the steady-state density of $\rho=\frac{1}{3}$ is purely
exponential.  We now recast Eqs.~\eqref{evol} as
\begin{align*}
\dot V_1 &= -5V_1+ 2\rho\\
\dot V_2 &= -6V_2+ 2\rho\\
\dot V_3 &= -7V_3-2V_1+ 2\rho\\
\dot V_4 &= -8V_4-2V_1-2V_2+ 2\rho\,,
\end{align*}
etc., which we can solve recursively to give
\begin{align}
\label{Vt}
\begin{split}
V_1 &= \tfrac{1}{15}\left(2-5e^{-3t}+3e^{-5t}\right)\\
V_2 &= \tfrac{1}{9}\left(1-e^{-3t}\right)^2\\
V_3 &= \tfrac{1}{35}\left(2-7e^{-5t}+5e^{-7t}\right)\\
V_4 &= \tfrac{1}{45}\left(1+4e^{-3t}-6e^{-5t}-5e^{-6t}+6e^{-8t}\right)\,,
\end{split}
\end{align}
etc., for an initially empty system.  Since each void density
approaches its steady-state value exponentially quickly, we now focus
on the steady state, where Eqs.~\eqref{evol} reduce to
\begin{equation}
\label{V:eq}
(k+4)V_k = 2\sum_{j\geq k-1}V_j \,.
\end{equation}
Introducing the cumulative distribution $F_k \equiv \sum_{j\geq k}V_j$, \eqref{V:eq} becomes
\begin{equation}
\label{F:eq}
F_{k+1}-F_{k+2} = \frac{2}{k+5}\,F_{k}\,.
\end{equation}
The first two of Eqs.~\eqref{conditions} give
$F_0=F_1=\rho=\frac{1}{3}$; these serve as the initial conditions that
allow us to generate all the $F_k$ one by one: $F_2=\frac{1}{5}$,
$F_3=\frac{4}{45}$, $F_5=\frac{2}{63}$, etc.

To find the general solution of Eq.~\eqref{F:eq} we employ the
generating function
technique~\cite{wilf2005generatingfunctionology}. The factor
$(k+5)^{-1}$ on the right-hand side of \eqref{F:eq} suggests that it
is expedient to define the generating function as
\begin{equation*}
F(z) \equiv \sum_{k\geq 0} F_k z^{k+4}\,.
\end{equation*}
Multiplying Eq.~\eqref{F:eq} by $z^{k+5}$ and summing over all $k\geq 0$, we
transform the recurrence \eqref{F:eq} into the integral equation
\begin{equation}
\label{Fz:eq}
F(z)-\rho z^4-\frac{F(z)-\rho z^4-\rho z^5}{z}=2\int_0^z dw\,F(w)\,.
\end{equation}
We now define
\begin{equation*}
\Phi(z) = \int_0^z dw\,F(w) = \sum_{k\geq 0} \frac{F_k}{k+5}\, z^{k+5}\,,
\end{equation*}
and after some elementary manipulations, we may express \eqref{Fz:eq}
as the ordinary differential equation
\begin{equation}
\label{Phi:eq}
 (1-z)\,\frac{d\Phi}{dz} + 2z \Phi = \rho z^4\,.
\end{equation}

Integrating \eqref{Phi:eq} subject to $\Phi(0)=0$ yields 
\begin{align*}
\Phi = \rho\,(1-z)^2 e^{2z}\int_0^z dw \, \frac{w^4\,e^{-2w}}{(1-w)^3}
= \tfrac{1}{4}\,\rho\big[3(1-z)^2 e^{2z}-3 + 3 z^2 + 2 z^3\big]\,.
\end{align*}
Finally, we differentiate $\Phi$ to give the generating function
\begin{equation}
\label{F:sol}
F(z) = \tfrac{3}{2}\,\rho\,\big[z+ z^2 -z(1-z) e^{2z}\big]\,.
\end{equation}
We now expand $F(z)$ in a power series to extract the $F_k$:
\begin{equation*}
F_k = 2^{k+1}\,\frac{k+1}{(k+3)!}\,,
\end{equation*}
from which the density of voids of length $k$ is
\begin{equation}
\label{Vk:sol1}
V_k =F_k - F_{k+1} = 2^{k+1}\,\frac{k(k+3)}{(k+4)!}\,.
\end{equation}
The average void length $\langle k\rangle = \sum kV_k/\sum V_k=2$, which
accords both with $\rho=\frac{1}{3}$ and with the conditions
\eqref{conditions}.  Higher moments of the void length are less simple:
$\langle k^2\rangle =3e^2\!-\!17\approx 5.167$,
$\langle k^3\rangle =83\!-\!9e^2\approx 16.499$, etc.

A basic question about the steady state is: how many birds fly away
after each landing event?  According to our model definition, either
0, 1, or 2 birds can fly away when a bird lands.  The probabilities
$q_n$ that $n\leq 2$ birds fly away after each landing event satisfy
the sum rules
\begin{align*}
  q_0+q_1+q_2=1\,,\qquad
  q_1+2q_2=1\,.
\end{align*}
The first equation imposes normalization, while the second equation states
that in the steady state one bird flies away, on average, after each landing
event.  These lead to $q_0=q_2$.  The probabilities $q_n$ are determined by
\begin{align*}
  \label{qn}
  q_0&=\sum_{k\geq 3}\frac{(\!k-\!2) V_k}{1\!-\!\rho}\qquad
  q_1= \sum_{k\geq 2} \frac{2 V_k}{1\!-\!\rho}\qquad   q_2= \frac{V_1}{1\!-\!\rho}\,.
\end{align*}
The first term accounts for a bird landing in the interior of a gap of length
$k\geq 3$ so that no bird leaves.  The second term accounts for a bird
landing at either end of a gap of length $k>2$ so that a single bird leaves.
The last term account for a bird landing in a vacancy between two birds so
that both these birds leave.  The normalization factor $(1-\rho)^{-1}$ is the
probability to land on any vacancy.  Using $\rho=\frac{1}{3}$ and
Eq.~\eqref{Vk:sol1}, we find $q_0=q_2=\frac{1}{5}$, $q_1=\frac{3}{5}$.

We can also readily extend our approach to treat the situation in
which all birds within a range $b>1$ fly away when a bird lands on an
unoccupied site.  While the qualitative features of this
generalization are the same as that for the case $b=1$ given above,
some quantitative differences arise.  The solution for general $b>1$
is given in \ref{app:b>1}.

\section{The pair correlation function}
\label{ap:CF}

The spatial distribution of birds may be characterized by the pair
correlation function $C_j \equiv \langle n_0 n_j\rangle$, where $n_j$
is the occupancy indicator function at site $j$.  That is, $n_j=0$ if
site $j$ is empty and $n_j=1$ if $j$ is occupied. If the locations of
the birds are spatially uncorrelated, then
$\langle \langle n_0 n_j\rangle= \langle n_0 \rangle\langle n_j
\rangle$.  This implies that the connected correlation function,
$\mathcal{C}_j= \langle n_0 n_j\rangle - \langle n_0\rangle \langle
n_j\rangle$ would equal zero.  Our calculations below seem to suggest
that this is the case.  The connected correlation functions
$\mathcal{C}_1$ and $\mathcal{C}_2$ are non zero, while we show that
$\mathcal{C}_4$, and $\mathcal{C}_5$ are zero.  These calculations
become quite tedious for $\mathcal{C}_4$ and $\mathcal{C}_5$ and we
can only conjecture that $\mathcal{C}_j=0$ for $j>5$.

The steady-state pair correlation function $C_j$ for $j\leq 3$ can be
deduced directly from our results for the density and the void
densities.  Indeed,
$C_0 = \langle n_0^2\rangle=\langle n_0\rangle=\frac{1}{3}$, while
$C_1=V_0$, $C_2 = V_1$ and $C_3 = V_2$, from which
\begin{equation}
\label{C0123}
C_0 = \tfrac{1}{3}\,, \quad C_1 = 0\,, \quad C_2 = \tfrac{2}{15}\,, \quad   C_3 = \tfrac{1}{9}\,.
\end{equation}

We now derive $C_4=C_5=\frac{1}{9}$.  As we show, determining these
correlation functions requires various multi-void distributions.  The
formal expressions for the first few correlation functions $C_j$, with
$j\geq 4$, are:
\begin{align*}
C_4 & = \text{Prob}\Big[\!\bullet\!\circ\!\bullet\!\circ\!\bullet\!\Big]+\text{Prob}\Big[\!\bullet\!\circ\!\circ\!\circ\!\bullet\!\Big]=V_{1,1} + V_3\\
C_5 & = \text{Prob}\Big[\!\bullet\!\circ\!\bullet\!\circ\!\circ\!\bullet\!\Big]+\text{Prob}\Big[\!\bullet\!\circ\!\circ\!\bullet\!\circ\!\bullet\!\Big] + \text{Prob}\Big[\!\bullet\!\circ\!\circ\!\circ\!\circ\!\bullet\!\Big]= 2 V_{1,2}+V_4\\
  C_6 & = \text{Prob}\Big[\!\bullet\!\circ\!\bullet\!\circ\!\bullet\!\circ\!\bullet\!\Big]
        + \text{Prob}\Big[\!\bullet\!\circ\!\circ\!\bullet\!\circ\!\circ\!\bullet\!\Big] 
        + \text{Prob}\Big[\!\bullet\!\circ\!\bullet\!\circ\!\circ\!\circ\!\bullet\!\Big]\\
        &\qquad\qquad+ \text{Prob}\Big[\!\bullet\!\circ\!\circ\!\circ\!\bullet\!\circ\!\bullet\!\Big] 
         + \text{Prob}\Big[\!\bullet\!\circ\!\circ\!\circ\!\circ\!\circ\!\bullet\!\Big]\\
        & = V_{1,1,1} + V_{2,2} + 2V_{1,3}+V_5\,,
\end{align*}
where
\begin{align*}
&V_k         \equiv \text{Prob}\Big[\bullet\underbrace{\circ\,\ldots\,\circ}_k\bullet\Big] \\
&V_{i,j}     \equiv \text{Prob}\Big[\bullet\underbrace{\circ\,\ldots\,\circ}_i\bullet\underbrace{\circ\,\ldots\,\circ}_j\bullet\Big]\\
&V_{i,j,k}   \equiv \text{Prob}\Big[\bullet\underbrace{\circ\,\ldots\,\circ}_i\bullet\underbrace{\circ\,\ldots\,\circ}_j\bullet\underbrace{\circ\,\ldots\,\circ}_k\bullet\Big]\,,
\end{align*}
denote the single-void, two-void, and three-void distributions.  The
subscripts on the multi-void distributions account for the number of
sites in the adjacent empty strings.

The void distributions $V_{i_1,\ldots,i_p}$ satisfy rate equations
that are natural extensions of the rate equation \eqref{evol} for
$V_k$.  Consider first the distribution $V_{i,j}$.  Using the same
reasoning as that given in Fig.~\ref{fig:processes} to write
Eq.~\eqref{evol}, the rate equation for $V_{i,j}$ is
\begin{align}
\label{Vij:eq}
\dot V_{i,j} &= -(2+i+j)V_{i,j} + \sum_{\ell\geq i+1}V_{\ell,j}+ \sum_{\ell\geq j+1}V_{i,\ell} \nonumber \\
                  &\qquad\qquad+ V_{i+j+1} +V_{i-1,j+1}+V_{i+1,j-1}+V_{i-1,j}+V_{i,j-1}\,,
\end{align}
subject to the boundary conditions
\begin{equation}
\label{V0-ij}
V_{i,0} = 0 = V_{0,j}\qquad (i, j\geq 0), 
\end{equation}
and the sum rules
\begin{equation}
\label{VVVV}
\sum_{\ell\geq 1}V_{\ell,j}=V_j, \qquad \sum_{\ell\geq 1}V_{i,\ell}=V_i\,.
\end{equation}

In the steady state, \eqref{Vij:eq} reduces to the recurrence
\begin{equation}
\label{Vij:rec}
(4+i+j)V_{i,j} = \sum_{\ell\geq i}V_{\ell,j}+ \sum_{\ell\geq j}V_{i,\ell} +V_{i+j+1} +V_{i-1,j+1}+V_{i+1,j-1}+V_{i-1,j}+V_{i,j-1}\,.
\end{equation}
Specializing \eqref{Vij:rec} and \eqref{VVVV} to $(i,j)=(1,1)$ and
additionally using \eqref{V0-ij} we obtain
\begin{equation}
\label{V11}
6V_{1,1}=2V_1+V_3\,.
\end{equation}
Recalling that $V_1=\frac{2}{15}$ and $V_3=\frac{2}{35}$ from
Eq.~\eqref{Vk:sol1}, we obtain $V_{1,1}=\frac{17}{315}$, which finally gives
$C_4=V_{1,1}+V_3=\frac{1}{9}$.  

Next, we specialize \eqref{Vij:rec} and \eqref{VVVV} to $(i,j)=(1,2)$,
from which we obtain
\begin{equation}
\label{V12}
6V_{1,2}=V_2+V_1+V_4
\end{equation}
Using the known results $V_1=\frac{2}{15}, V_2=\frac{1}{9}, V_4=\frac{1}{45}$ we obtain
$V_{1,2}=\frac{2}{45}$ and then $C_5=2V_{1,2}+V_4=\frac{1}{9}$.  

We mention that we can determine the full time dependence of the
low-order pair correlation functions.  The behaviors of $C_j$ with
$j=0,1,2$, and 3 follow directly from the relation between these
correlation functions and the appropriate void densities.  Namely,
$C_0(t)=\rho(t)$, $C_1(t)=V_0(t)$, $C_2(t) = V_1(t)$ and
$C_3(t) = V_2(t)$.  To derive $C_4(t)= V_{1,1}(t) + V_3(t)$
we must find $V_{1,1}(t)$.  From \eqref{Vij:eq} The rate equation for $V_{1,1}$ is
\begin{equation*}
\dot V_{1,1} = -6V_{1,1} + 2V_1+V_3\,,
\end{equation*}
with solution, for an initially empty system, 
\begin{equation}
\label{V11:sol}
V_{1,1} = \tfrac{1}{315}\left(17 - 70 e^{-3t}+63 e^{-5t} + 35 e^{-6t} - 45 e^{-7t}\right)\,.
\end{equation}
Using $C_4=V_{1,1} + V_3$ with $V_3(t)$ from \eqref{Vt} and
$V_{1,1}(t)$ from \eqref{V11:sol} we have
\begin{equation}
\label{C4}
 C_4  = V_{1,1} + V_3 = \tfrac{1}{9}\left(1-e^{-3t}\right)^2\\
\end{equation}

To derive \eqref{C5}, we must find $V_{1,2}(t)$.  Again from
\eqref{Vij:eq}, the rate equation for $V_{1,2}$ is
\begin{equation*}
\dot V_{1,2} = -6V_{1,2} + V_1+V_2+V_4\,,
\end{equation*}
whose solution is
\begin{equation}
\label{V12:sol}
V_{1,2} = \tfrac{1}{45}\left(2 - 7 e^{-3t}+3 e^{-5t} + 5 e^{-6t} - 3 e^{-8t}\right)\,.
\end{equation}
Using $C_5=2V_{1,2}+V_4$ with $V_4(t)$ from \eqref{Vt} and
$V_{1,2}(t)$ from \eqref{V12:sol} we thus find
\begin{equation}
   \label{C5}
   C_5  = 2V_{1,2}+V_4 = \tfrac{1}{9}\left(1-e^{-3t}\right)^2\,.
\end{equation}
It seems unlikely that we can determine of the correlation functions
$C_j$ for arbitrary $j$ via this straightforward, but laborious method.

\section{One-dimensional continuum}

A more natural scenario for the dynamics is that each birds can land
anywhere along a wire.  Within the RSA framework, the analogous
process is the famous R\'enyi car parking model~\cite{Renyi58} in which
fixed-length cars attempt to park anywhere along a one-dimensional
line until there are no gaps remaining that can accommodate a
car. Without loss of generality we set the interaction range between
birds equal to one.  Thus if a bird lands within a unit distance of
one (or two) birds, this bird (or these birds) immediately fly away.

\begin{figure}[ht]
\centerline{\subfigure[]{\raisebox{2mm}{\includegraphics[width=0.3\textwidth]{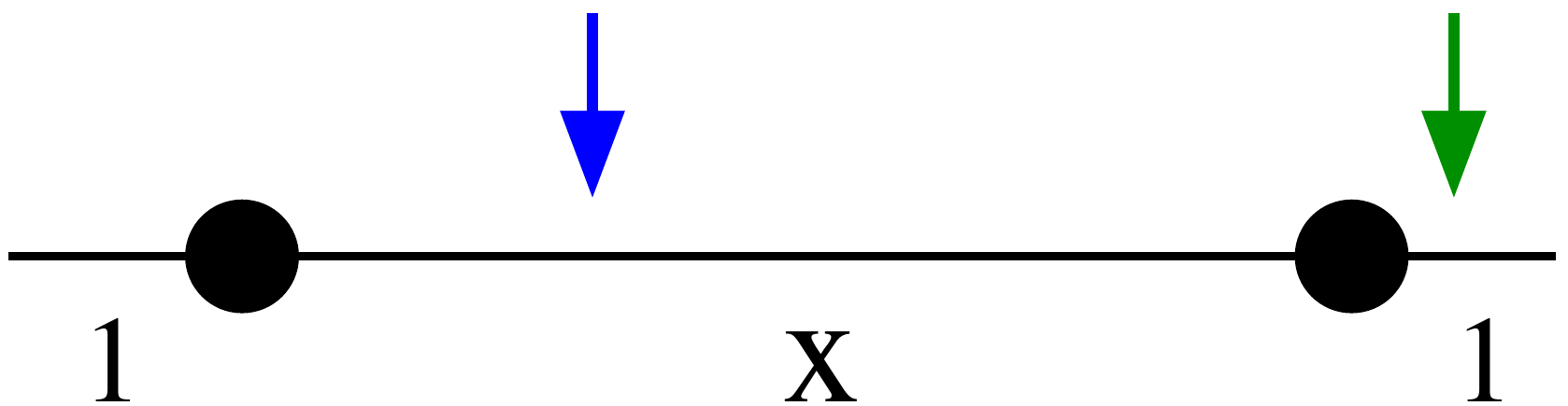}}}\qquad\qquad
\subfigure[]{\includegraphics[width=0.3\textwidth]{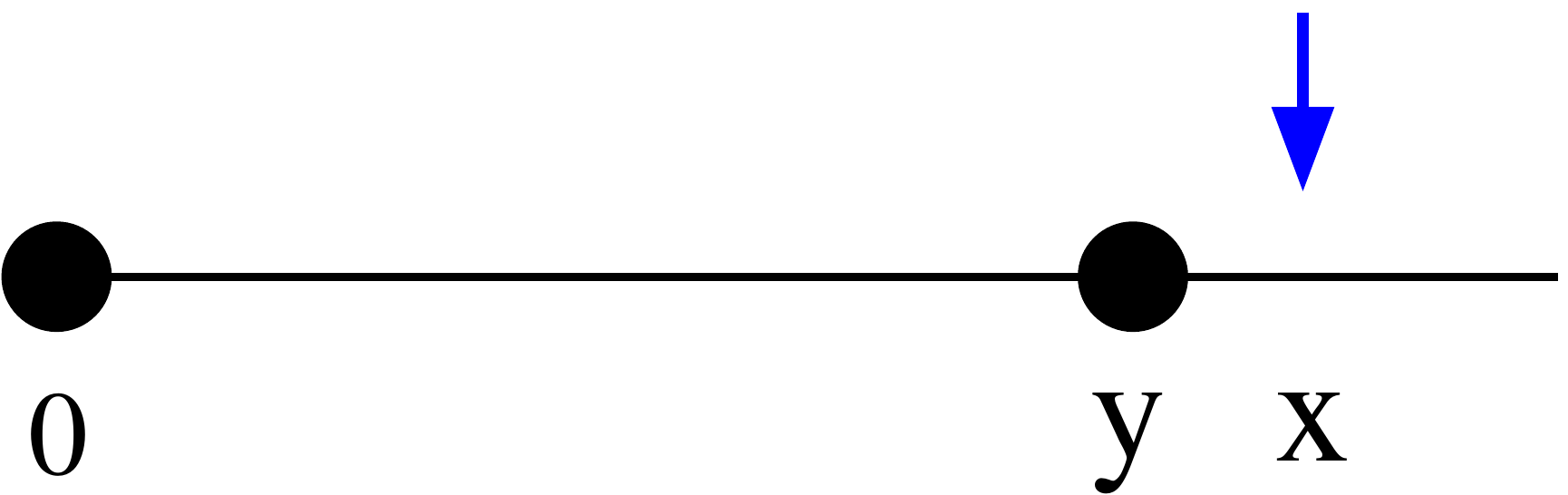}}}
\caption{Processes that contribute to changes in the void densities in
  Eq.~\ref{Vx12}.  (a) An $x$-void disappears if a bird lands anywhere
  inside the void (blue arrow) or within a unit distance of either
  bird outside the void (green arrow), (b) An $x$-void is created when
  a new bird lands a distance $x$ from an existing bird. Another bird
  may be anywhere in the range $[1,\infty]$ for $1<x<2$ or in the
  range $[x-1,\infty]$ for $x>2$. }
\label{fig:processes-cont}  
\end{figure}

Instead of voids of integer length, the basic dynamical variable is
$V(x)$, the density of voids of length $x$.  Following the same
reasoning as that which led to Eq.~\eqref{evol}, the evolution
equation for the void distribution is now (see also Fig.~\ref{fig:processes-cont})
\begin{align}
\label{Vx12}
  \dot V(x,t) =   -(2+x)V(x,t)+
\begin{cases}
{\displaystyle 2\int_{1}^\infty dy\,V(y,t)}&\qquad\qquad 1<x<2\,,\\[5mm]
{\displaystyle 2\int_{x-1}^\infty dy\,V(y,t)}&\qquad\qquad x>2\,.
\end{cases}
\end{align}
In close analogy with Eq.~\eqref{conditions}, the void distribution
$V(x)$ must now satisfy the sum rules: (a) $V(x) = 0$ for $x<1$, (b)
the density of birds is $\rho = \int_1^\infty dx\,V(x)$, and (c)
$\int_1^\infty dx\,x\, V(x)=1$.  As a result of condition (b), the first
of Eqs.~\eqref{Vx12} can be re-expressed as
$\dot V(x,t)= -(2+x) V(x,t)+2\rho(t)$.

Integrating \eqref{Vx12} over all $x$, the density
\begin{equation*}
\rho(t) = \int_1^2 dx\,V(x,t)+\int_2^\infty dx\,V(x,t)
\end{equation*}
obeys the rate equation $\dot \rho=1-2\rho$.  For an initially empty
system, the solution is simply $\rho=\frac{1}{2}(1-e^{-2t})$.  We now
use this result $\rho$ to solve $\dot V(x,t)= -(2+x) V(x,t)+2\rho(t)$
in the range $1<x<2$ to give
\begin{equation}
\label{Vx12:sol}
V(x,t)= \frac{1-e^{-(2+x)t}}{2+x}-\frac{e^{-2t}- e^{-(2+x)t}}{x}\,.
\end{equation}
Using $\rho = \frac{1}{2}(1-e^{-2t})$ in the second of \eqref{Vx12}, we
may rewrite this equation as
\begin{eqnarray}
\label{Vx2-rec}
\dot V(x,t) = -(2+x)V(x,t)-2\int_1^{x-1}dy\,V(y,t)+1-e^{-2t}\,.
\end{eqnarray}
We now substitute the solution for $V(x,t)$ in the range $1<x<2$ in
Eq.~\eqref{Vx2-rec} to solve this equation in the interval $2<x<3$.
Continuing this procedure we can recursively solve \eqref{Vx2-rec} for
each interval $n<x<n+1$ using the previously determined solutions for
$x<n$.  While this procedure is straightforward in principle, it
quickly becomes tedious as $x$ increases.

To obtain the large-$x$ behavior of the void distribution, we first
rely on the fact that the approach to the steady state again occurs
exponentially quickly.  Thus we henceforth focus on the steady-state
properties of the continuum case.  In this case, the void density is
determined by
\begin{equation}
  \label{V:cont}
  (2+x)V(x)=
  \begin{cases}
{\displaystyle 2\int_{1}^\infty dy\,V(y)}&\qquad\qquad x<1\,,\\[5mm]
{\displaystyle 2\int_{x-1}^\infty dy\,V(y)}&\qquad\qquad 1<x<2\,.
\end{cases}
\end{equation}
To solve Eq.~\eqref{V:cont}, we introduce the Laplace transform
$\widehat{V}(s)\equiv \int_1^\infty dx\,e^{-xs}\, V(x)$.  Then the
Laplace transform of the left-hand side of Eq.~\eqref{V:cont} is
\begin{equation*}
\int_1^\infty dx\,e^{-xs}(2+x)V(x)=2\widehat{V}-\frac{d\widehat{V}}{ds}\,.
\end{equation*}
The Laplace transform of the right-hand side of the first of
\eqref{V:cont} is, after accounting for the constraint $1<x<2$,
\begin{align*}
2\rho \int_1^2 dx\,e^{-xs} = \frac{2\rho}{s}\,(e^{-s}-e^{-2s})\,.
\end{align*}
Similarly, the Laplace transform of the right-hand side of the second
of \eqref{V:cont} is, after accounting for the constraint $x>2$,
\begin{align*}
  \int_2^\infty \! dx\,e^{-xs} \int_{x-1}^\infty dy\,V(y)
  =  \int_{1}^\infty \!\!\! dy\,V(y) \int_2^{y+1}\!\! dx\,e^{-xs} 
= \frac{e^{-s}}{s}\,\left(\rho e^{-s}-\widehat{V}\right)\,.
\end{align*}

Using these results, the Laplace transform satisfies
\begin{equation}
 \label{Vs:eq}
2\left(1+s^{-1}e^{-s}\right)\widehat{V}-\frac{d\widehat{V}}{ds}=2\rho s^{-1}e^{-s}\,.
\end{equation}
Integrating \eqref{Vs:eq} and using the steady-state density
$\rho=\frac{1}{2}$ yields
\begin{equation}
\label{V:final}
\widehat{V}(s) = \frac{1}{2}-\mathcal{E}(s)\int_s^\infty
\frac{d\sigma}{\mathcal{E}(\sigma)}\,.
\end{equation}
where we define $\mathcal{E}(s)\equiv e^{2s-2 E_1(s)}$ and $E_1$ is the
exponential integral~\cite{abramowitz1964handbook}
\begin{equation*}
E_1(s) = \int_s^\infty \frac{d\sigma}{\sigma}\,e^{-\sigma}\,.
\end{equation*}

The large-$x$ behavior of $V(x)$ is in principle encoded in the Laplace
transform $\widehat{V}(s)$.  However, it is easier to extract the asymptotic
behavior from the derivative of Eq.~\eqref{V:cont}, namely, from
\begin{equation}
\label{V:diff}
[(2+x)V(x)]'= -2V(x-1)\,,
\end{equation}
where the prime denotes differentiation with respect to $x$.  We will find
that $V(x)$ decays super-exponentially with $x$ for large $x$.  Thus a Taylor
expansion of $V(x)$ is not justified.  Instead we seek a solution of the form
$V(x) = e^{-w(x)}$, where it is justifiable to expand $w(x-1)$ as
$w(x) - w'(x)$.  Doing so in Eq.~\eqref{V:diff} gives $x w' = 2 e^{w'}$ to
leading order.  The solution to this equation is
\begin{equation}
w= x[\ln x + \ln(\ln x)-1-\ln 2] +\ldots\,.
\end{equation}
Thus the void density $V(x)=e^{-w(x)}$ exhibits essentially a
factorial (faster than exponential) decay.

\begin{figure}[ht]
\begin{center}
\includegraphics[width=0.45\textwidth]{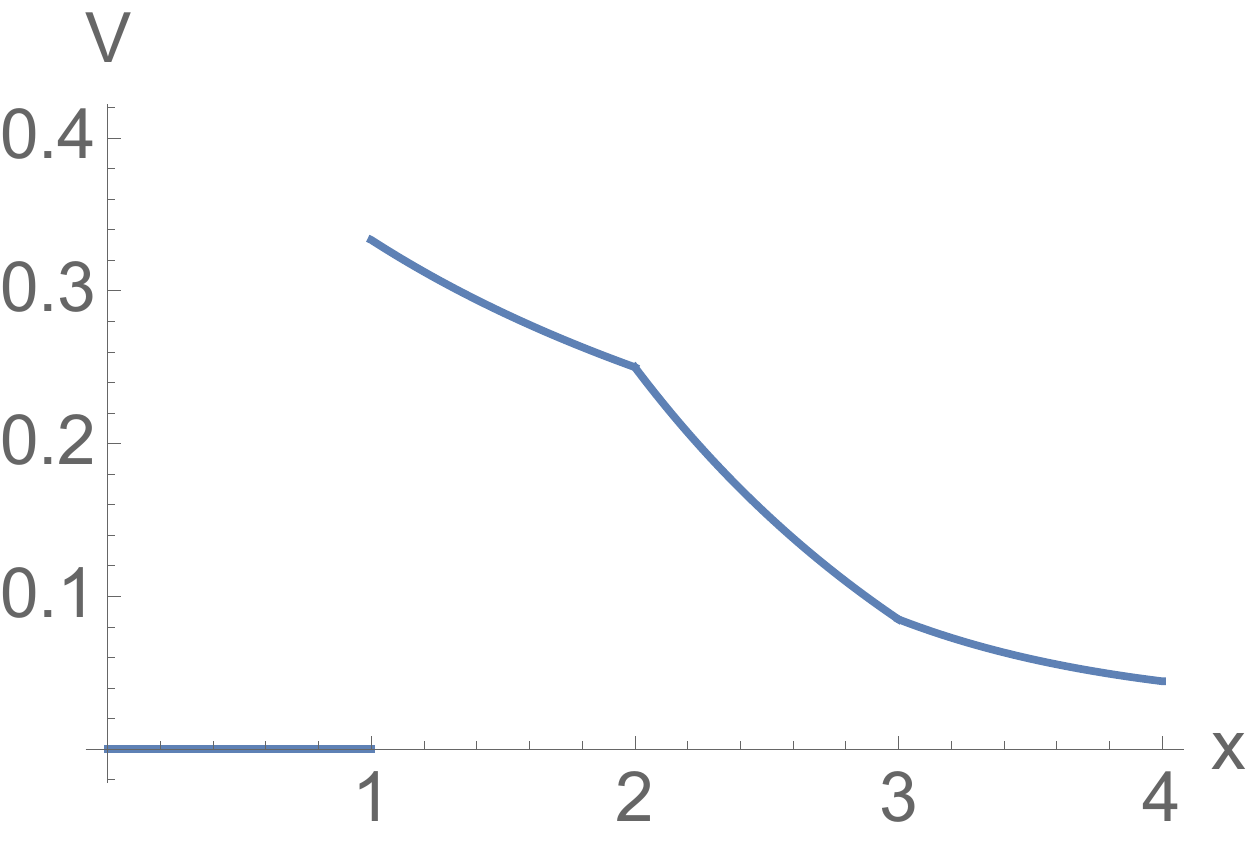}
\caption{The void density $V(x)$ for $0<x<4$, showing the jump at $x=1$ and
  singularities in the first derivative at $x=2$ and $x=3$.}
\label{fig:V1234}
  \end{center}
\end{figure}

We can use the result $\rho = \frac{1}{2}$ to directly find $V(x)$ in
the successive intervals $1<x<2$, $2<x<3$, etc., from \eqref{V:cont}
without recourse to the Laplace transform method.  From the first of
\eqref{V:cont}, we obtain
\begin{align}
V(x)=
\begin{cases}
\label{V:1-2}
\displaystyle{ \frac{1}{2+x}}& 1<x<2\,, \\[8mm]
\displaystyle{ \frac{\left[1-2\ln\left((x+1)/3\right)\right]}{2+x}}& 2<x<3\,.
\end{cases}
\end{align}
For $x>2$, we recast the first of Eqs.~\eqref{V:cont} into
\begin{equation}
(2+x)V(x) = 1 - 2\int_1^{x-1} dy\,V(y)\,,
\end{equation}
from which the density, for $3<x<4$, is
\begin{align*}
(2+x)V(x)  = 1-\ln(4/9)  -2\text{Li}_2(-3)+2\text{Li}_2(-x)
 -(1+ 2\ln 3 - 2\ln x)\ln(1+x)\,,
\end{align*}
where $\text{Li}_2(-x) = \sum_{j\geq 1}(-x)^j/j^2$ is the dilogarithm
function~\cite{abramowitz1964handbook}.  One may continue this
iterative procedure to obtain explicit expressions for $V(x)$ for
$n<x<n+1$ for positive integer $n$.  These calculations quickly become
tedious, so we do not extend them beyond $x=4$.  The resulting
function $V(x)$ is singular (Fig.~\ref{fig:V1234}) with a slope
discontinuity at every integer $x\geq 2$; thus inversion of the
Laplace transform~\eqref{V:final} in terms of a compact formula is
also not possible.  The main features of the void distribution $V(x)$
is that it is a piecewise smooth function, with increasingly
cumbersome expressions for $V(x)$ for $n<x<n+1$, and which decays as
$x^{-x}$ for large $x$.

In analogy to the argument that led to the probabilities $q_n$ for $n$
birds to fly away at each landing event in the lattice model, in the
1d continuum version the corresponding probabilities are
\begin{align}
  \label{q-cont}
  \begin{split}
q_0 & = \int_{2}^\infty dx\,(x-2)V(x)\\
q_1 & =2\int_{2}^\infty dx\,V(x) + 2\int_1^2 dx\,(x-1)V(x)\\
q_2 & = \int_1^2 dx\,(2-x)V(x)
\end{split}
\end{align}
Using $\rho = \frac{1}{2}$ and \eqref{V:1-2} we find
$q_0  = q_2 = 4\ln(4/3)-1\approx 0.151$ and $q_1  = 3-8\ln(4/3)\approx 0.699$.

\section{Higher dimensions}

Our PB model naturally extends to the realistic situation of multiple
wires, as in Fig.~\ref{fig:wire}, and to higher dimensions.  On
hyper-cubic lattices $\mathbb{Z}^d$, we posit that all resting birds
that are one lattice spacing from the newly arriving bird fly away.
In the continuum $\mathbb{R}^d$, all resting birds within a unit
distance of the newly arriving bird fly away.  Simulations of the PB
model on various substrates show that an initially empty system
quickly reaches a steady state, and the steady-state densities are
$\rho \approx \frac{1}{5}$ and $\rho \approx \frac{1}{7}$,
respectively, for the square and cubic lattices.  These results lead
to conjectural steady-state densities on $d$-dimensional hyper-cubic
lattices
\begin{align}
  \label{gen-d}
  \rho = \frac{1}{2d+1}\,.
\end{align}               
The derivation of this result is left to future work. 

It is also instructive to construct a mean-field theory for the steady-state
density of the PB model on hypercubic lattices.  This theory is based on
neglecting correlations in the spatial positions of the birds.  In this
approximation, the density of birds on a $d$-dimensional hypercubic lattice
obeys the rate equation
\begin{equation}
\label{rho:MFT}
\frac{d\rho}{dt}=-(1-\rho)\sum_{n=0}^{2d}(n-1)\binom{2d}{n}(1-\rho)^{2d-n}\rho^{n}\,.
\end{equation}
The $n=0$ term in this sum is positive corresponds to the case where the bird
lands on an empty site and all neighbors of this site are also empty, so that
no birds fly away and $\rho$ increases.  The terms with $n\geq 1$ are
non-negative and correspond to the situations where at least one resting bird
flies away when the bird lands.  Equation~\eqref{rho:MFT} simplifies to
$\frac{d\rho}{dt}=(1-\rho)(1-2 \rho\, d)$. This gives the steady-state
density $\rho = \frac{1}{2d}$, which approaches the exact steady state \eqref{gen-d}
in the limit $d\to\infty$.  From this same mean-field argument, the
probabilities $q_n$ for $n$ birds to fly away, with $0\leq n\leq 2d$, after
each landing event is
\begin{align}
  q_n =\binom{2d}{n}(1-\rho)^{2d-n}\rho^n \,.
\end{align}
Using the mean-field steady-state density $\rho = \frac{1}{2d}$, the above
expression reduces to $q_n = e^{-1}/n!$ as $d\to\infty$.  This is a rapidly
decaying distribution, so that the average size of the ``avalanche'' that is
nucleated when a bird lands is small: $\langle n\rangle = 1-e^{-1}$.

\section{Concluding comments}

Our PB model is inspired by natural observations and seamlessly leads
to a simple non-equilibrium statistical physics model of competing
adsorption/desorption.  We solved for the time-dependent and
steady-state properties of the model analytically.  An appealing
challenge is to determine the steady-state properties of the PB model
in general dimensions, both on lattices and on a continuum.  Another
potentially fruitful direction is to extend to realistic longer-range
interactions between birds.  In such a scenario, when a bird lands, it
may drive a large groups of birds to fly away.  This type of slow
driving and sudden large ``avalanches'' is reminiscent of the size of
fires in self-organized forest fire
models~\cite{drossel1992self,drossel1993exact}, as well as the size of mass
rearrangements in the random organization
model~\cite{wilken2021random,corte2008random}.

\section*{Acknowledgments}
We thank O. Adelman, R. Dandekar, D. Dhar, D. S\'{e}nizergues and
N. Smith for useful correspondence.  SR gratefully acknowledges
partial financial support from NSF grant DMR-1910736.

\appendix

\section{Birds with Interaction Range $b>1$}
\label{app:b>1}

We outline some basic steady-state properties of our PB model on a discrete
one-dimensional lattice in which, after each landing event, all birds that
are within a distance $b$ of the incident bird fly away.  While the solution
for the generating function can again be obtained by following the steps from
Eqs.~\eqref{V:eq}--\eqref{F:sol}, this calculation becomes cumbersome as $b$
increases.  However, the steady-state density $\rho=1/(2b+1)$ can be
extracted fairly easily without the complete solution for the void
distribution.

Let us first treat the case $b=2$; the extension for $b>2$ then readily
follows. In the steady state, the generalization of Eq.~\eqref{V:eq} for the
void densities $V_k$ is 
\begin{equation}
\label{V2:eq}
(k+6)V_k = 2\sum_{j\geq k-2}V_j = 2 F_{k-2}\,.
\end{equation}
We use the initial conditions $V_0 = V_1 =0$, as well as
$\rho=\sum_{k\geq 0} V_k$ to solve \eqref{V2:eq} recursively and obtain
\begin{equation}
\label{V234}
V_2 = \tfrac{1}{4}\rho, \qquad V_3 = \tfrac{2}{9}\rho, \qquad V_4 = \tfrac{1}{5}\rho, \qquad V_5 = \tfrac{3}{22}\rho\,,
\end{equation}
etc.  By using the generating function technique, we can fix $\rho$ and then
determine $V_k$ for arbitrary $k$.  However, if we merely want to find the
steady-state density, we adopt the following approach.  We first rewrite \eqref{V2:eq}
as
\begin{equation}
\label{F2:eq}
(k+8)[F_{k+2}-F_{k+3}] = 2 F_{k}\,,
\end{equation}
and then sum over all $k\geq 0$ to yield
\begin{equation}
\label{F2:sum}
7F_2+\sum_{k\geq 2}F_k = 2 \sum_{k\geq 0}F_k\,.
\end{equation}

The initial conditions $V_0 = V_1 =0$ leads to $F_0 = F_1 = F_2 = \rho$,
which then allows us to reduce \eqref{F2:sum} to
\begin{equation}
\label{F2:density}
5\rho = \sum_{k\geq 0}F_k\,.
\end{equation}
Using the normalization condition
$\sum_{k\geq 0}(k+1)V_k  = \sum_{k\geq 0}F_k =1$
we arrive at the basic result
\begin{equation}
\label{2:density}
\rho = \frac{1}{5}\,.
\end{equation}

We now determine the probabilities $q_n$ that $n$ birds fly away after each
landing event.  First note that $q_2$ is given by
\begin{equation}
\label{p2:2}
q_2=\tfrac{5}{4}\left(2V_2+V_3\right)\,.
\end{equation}
The factor $\frac{5}{4}$ accounts for that fact that the fraction of
successful landing events in the steady state is $\frac{4}{5}$. The term
$2V_2$ accounts for the 2 landing spots inside a vacancy of length 2 that
leads to two birds flying away, while the term $V_3$ accounts for the fact
that the landing must be at the center of a gap of length 3 to trigger two
departures. Using \eqref{V234} and \eqref{2:density}, the remaining probabilities
$q_n$ are
\begin{equation}
\label{p012:2}
q_0=q_2=\frac{13}{72}\,, \qquad q_1=\frac{23}{36}\,.
\end{equation}

For the case of arbitrary $b$. The analog of Eq.~\eqref{F2:eq} is
\begin{equation}
\label{Fb:eq}
(k+3b+2)[F_{k+b}-F_{k+b+1}] = 2 F_{k}\,.
\end{equation}
Summing over all $k\geq 0$ we obtain 
\begin{equation}
\label{Fb:sum}
(3b+1)F_b+\sum_{k\geq b}F_k = 2 \sum_{k\geq 0}F_k
\end{equation}
The initial condition $F_0 = F_1= \ldots = F_{b} = \rho$ yields
$\sum_{k\geq b} F_k = \sum_{k\geq 0} F_k +b\rho$.  Using this in
\eqref{Fb:sum}, we obtain 
\begin{equation}
\label{Fb:density}
(2b+1)\rho = \sum_{k\geq 0}F_k\,.
\end{equation}
Now using the normalization condition $\sum_{k\geq 0}F_k =1$, the
steady-state density is
\begin{equation}
\label{rho-b}
  \rho=\frac{1}{2b+1}\,.
\end{equation}
From \eqref{Fb:eq} and \eqref{rho-b}, and using the initial condition
$F_j=\rho$ for $j\leq b$ as well as the definition of $V_k$ in terms of
$F_k$, we find
\begin{equation}
\label{Vj:b}
V_{b+j} = \frac{2}{2b+1}\,\frac{1}{3b+2+j}\,.
\end{equation}

\begin{figure}[ht]
\begin{center}
\includegraphics[width=0.5\textwidth]{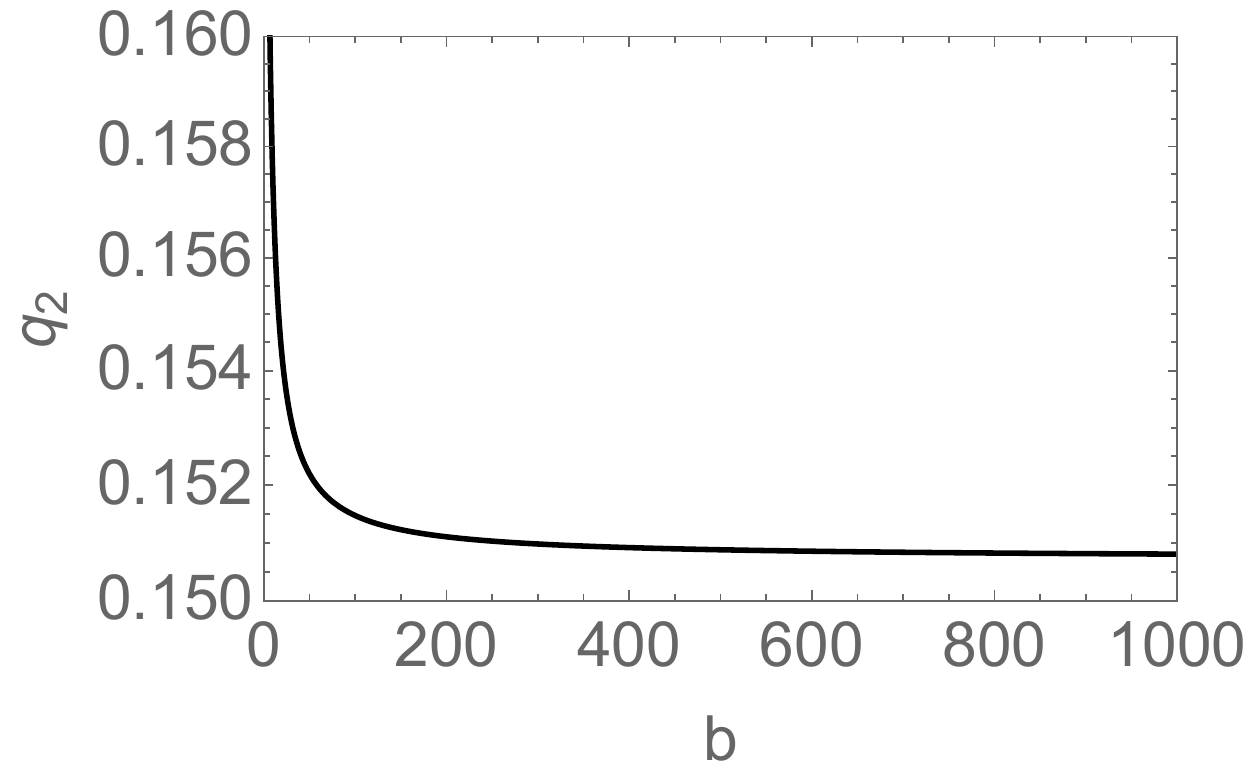}
\caption{The probability $q_2$ as a function of $b$ for $b\leq 1000$.}
\label{fig:q2}
  \end{center}
\end{figure}

Let us now determine the probabilities $q_n$ for arbitrary $b$.  The
generalization of \eqref{p2:2} is
\begin{equation}
\label{p2:b}
q_2=\frac{2b+1}{2b}\sum_{j=0}^b (b-j)V_{b+j}\,.
\end{equation}
The meaning of each term in the sum is the same as the two terms in
Eq.~\eqref{p2:2}: we are counting the number of ways that a bird can land
within a gap of length $b+j$ such that exactly two birds fly away.
Substituting in \eqref{Vj:b} into \eqref{p2:b} and computing the sum, we
obtain
\begin{equation}
\label{p02:b}
q_2=2(2+b^{-1})(H_{4b+2}-H_{3b+1})-1-b^{-1}\,,
\end{equation}
where $H_n=\sum_{1\leq j\leq n} j^{-1}$ is the $n^{\rm th}$ harmonic number.
Again, $q_0=q_2$ and $q_1$ is fixed by normalization, $q_1=1-2q_2$.  For
$b\to\infty$, $q_2\to 4\ln(4/3) -1\approx 0.15073$, which reproduces the
continuum result of Eq.~\eqref{q-cont}, as it must.  The dependence of $q_2$
on $b$ is shown in Fig.~\ref{fig:q2}.

Now we extend the above result to find the time-dependent behavior.
For general $b>1$, the void densities $V_k$ with $k\geq b$ evolve
according to
\begin{equation}
\label{Vk-b:eq}
\dot V_k = -(2b + 2+k)V_k+2\sum_{\ell\geq k-b} V_\ell\,,
\end{equation}
subject to the constraint that $V_0 = \ldots=V_{b-1} = 0$.  Summing
Eqs.~\eqref{Vk-b:eq} over $k\geq b$ and using the above constraint, as
well as Eqs.~\eqref{conditions}, we obtain the simple equation for the
density
\begin{equation*}
\dot\rho = 1 - (2b+1)\rho\,,
\end{equation*}
from which 
\begin{equation}
\label{rho:sol-b}
\rho(t) = \frac{1-e^{-(2b+1)t}}{2b+1}\,.
\end{equation}

The first non-trivial void density $V_b$ satisfies
\begin{equation}
\label{Vb:eq}
\dot V_b(t) = -(3b + 2)V_b+2\rho\,,
\end{equation}
from which
\begin{align}
\label{Vb:sol}
  V_b= \frac{2}{(2b+1)(3b+2)}- \frac{2\;e^{-(2b+1)t}}{(b+1)(2b+1)}
  + \frac{2\;e^{-(3b+2)t}}{(b+1)(3b+2)}\,.
\end{align}
The density $V_{b+1}$ satisfies
\begin{equation}
\label{Vb1:eq}
\dot V_{b+1} = -(3b + 3)V_{b+1}+2\rho\,,
\end{equation}
from which
\begin{align}
\label{Vb1:sol}
  V_{b+1}(t) = \frac{2}{(2b+1)(3b+3)}
  - \frac{2\;e^{-(2b+1)t}}{(b+2)(2b+1)}
  + \frac{2\;e^{-(3b+3)t}}{(b+2)(3b+3)}\,.
\end{align}

When $b\leq k\leq 2b$, the rate equation for $V_k$ has a form
\begin{equation}
\label{Vbj:eq}
\dot V_k = -(2b + 2+k)V_{b+j}+2\rho
\end{equation}
similar to \eqref{Vb:eq} and \eqref{Vb1:eq}. Solving \eqref{Vbj:eq} yields 
\begin{align}
\label{Vbj:sol}
V_k = \frac{2}{(2b+1)(2b+2+k)}- \frac{2\;e^{-(2b+1)t}}{(k+1)(2b+1)}
+ \frac{2\,e^{-(2b+2+k)t}}{(k+1)(2b+2+k)}
\end{align}
for $b\leq k\leq 2b$. 

\section{Higher-Order Correlation Functions}

The pattern in the equations for $C_4$, $C_5$, and $C_6$ generalizes
in straightforward way and we merely write the equations for the next
three correlation functions in the steady state:
\begin{align*}
C_7 & = 2V_{1,1,2} + V_{1,2,1}+2V_{1,4} + 2V_{2,3}+V_6\\
C_8 & = V_{1,1,1,1} + 2V_{1,1,3} + V_{1,3,1}
+2V_{1,5}+2V_{2,4} + V_{3,3}+V_7\\
C_9 & = 2V_{1,1,1,2} +2V_{1,1,2,1}  + 2V_{1,1,4} + V_{1,4,1}
 + 2V_{2,1,3} +2V_{1,2,3}+2V_{1,3,2} +V_{2,2,2}\\
&\qquad\qquad +2V_{1,6}+2V_{2,5} + 2V_{3,4}+V_8\,.
\end{align*}

\bibliographystyle{iopart-num}

\providecommand{\newblock}{}

\end{document}